\documentclass[twoside,a4paper,11pt]{sea9}
\usepackage{graphicx}
\usepackage{hyperref}
\usepackage{movie15}
\newcommand{\aap}{    {\it Astron. Astrophys.}}

\newcommand{\apjl}{   {\it Astrophys. J. Lett.}}

\begin{document}
\pagenumbering{arabic}
\pagestyle{myheadings}
\thispagestyle{empty}
{\flushleft\includegraphics[width=\textwidth,bb=58 650 590 680]{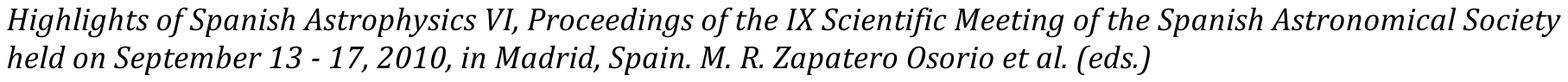}}
\vspace*{0.2cm}
\begin{flushleft}
{\bf {\LARGE
%
Stellar pulsations of solar-like oscillators with CoRoT and {\it Kepler}.
%
}\\
\vspace*{1cm}
%
R.A. Garc\'\i a$^{1}$
%
}\\
\vspace*{0.5cm}
%
$^{1}$
Laboratoire AIM, CEA/DSM-CNRS-Universit\'e Paris Diderot; CEA, IRFU, SAp, Centre de Saclay, 91191, Gif-sur-Yvette, France
%
\end{flushleft}
%
\markboth{
Solar-like oscillators with CoRoT and {\it Kepler}
}{ 
%
R.A. Garc\'\i a
%
}
\thispagestyle{empty}
\vspace*{0.4cm}
\begin{minipage}[l]{0.09\textwidth}
\ 
\end{minipage}
\begin{minipage}[r]{0.9\textwidth}
\vspace{1cm}
\section*{Abstract}{\small
%
Today, asteroseismology is entering in its golden age thanks to the observations provided by the CoRoT and {\it Kepler} space missions. In particular, we will make significant progresses in the understanding of the structure and evolution of solar-like oscillating stars. These stars have acoustic modes stochastically excited by the near-surface convection. Thanks to the observations already provided by these two missions, we have detected several hundred of stars showing solar-like oscillations in the main sequence and several thousands in the red-giant branch.
Here, I give an overview of the present status of the most important results obtained from both missions for stellar physics and the potential use of asteroseismology to characterize stars harboring planets.
%
\normalsize}
\end{minipage}
%
%
%
\section{Introduction \label{intro}}

For a long time, investigations of stellar interiors have been restricted to theoretical studies only constrained by observations of their global properties and external characteristics. However, in the last few decades the field has been revolutionized by the ability to perform seismic investigations of the internal properties of stars. Not surprisingly, it started by the Sun, where helioseismology (the seismic study of the Sun, e.g. Chistensen-Dalsgaard \cite{JCD2002}; Gough et al. \cite{1996Sci...272.1296G}) is yielding information rivaling what it can be inferred about the Earth's interior from geoseismology. 

The last few years have witnessed the advent of asteroseismology (the seismic study of stars, e.g. Bedding \& Kjeldsen \cite{2003PASA...20..203B}), thanks to a dramatic development of new observing facilities providing the first reliable results on the interiors of distant stars similar to the Sun. The coming years will see a huge development in this field.

Helio- and asteroseismology provide unique tools to infer the fundamental stellar properties (e.g., mass, radiusÉ) and to probe the internal conditions inside the Sun and stars (e.g. Stello et al. \cite{2009ApJ...700.1589S}). Today, asteroseismology also provides invaluable information to other scientific communities. As an example, it can give a good estimation of the masses, radii and ages of the stars hosting planets (Moya et al. \cite{2010MNRAS.406..566M}; Christensen-Dalsgaard et al. \cite{2010ApJ...713L.164C}; Gaulme et al. \cite{Gaulme2010}), that is a key-information to understand the formation of these planetary systems and their evolution, and also constrain the habitable zones of surrounding planets due to the stellar magnetic activity (e.g. Mosser et al. \cite{2005Mosser,2009Mosser}; Karoff et al. \cite{2009MNRAS.399..914K}; Mathur et al. \cite{2010A&A...518A..53M}; Garc\'\i a et al. \cite{2010Sci...329.1032G}; Metcalfe et al. \cite{2010ApJ...723L.213M}). This research leads to the testing and revision of our theories of stellar structure, dynamical processes, and evolution. Helio- and asteroseismology are today in a blooming phase both in scope and in magnitude. Helioseismology has shown the way to asteroseismology, which is reaching its maturity thanks to the CNES CoRoT satellite (Baglin et al. \cite{Bag2001}), the very promising NASA's {\it Kepler} spacecraft (Borucki et al. \cite{2010Sci...327..977B}; Koch et al. \cite{2010ApJ...713L..79K}) and the understudy ESA's PLATO mission (Catala et al. \cite{2008JPhCS.118a2040C}). It is important to remember the pioneers of this research done thanks to some episodic ground-based campaigns and some solar-like oscillating stars observed from space using both, the American satellite WIRE (Wide-Field Infrared Explorer, Buzasi et al. \cite{Buzasi2000}) and the Canadian MOST (Microvariability and Oscillations of Stars, Mathews 1998 \cite{mat98}).

\section{Helio and Astero-seismology}

\begin{figure}
\center
\includegraphics[width = 0.60\textwidth]{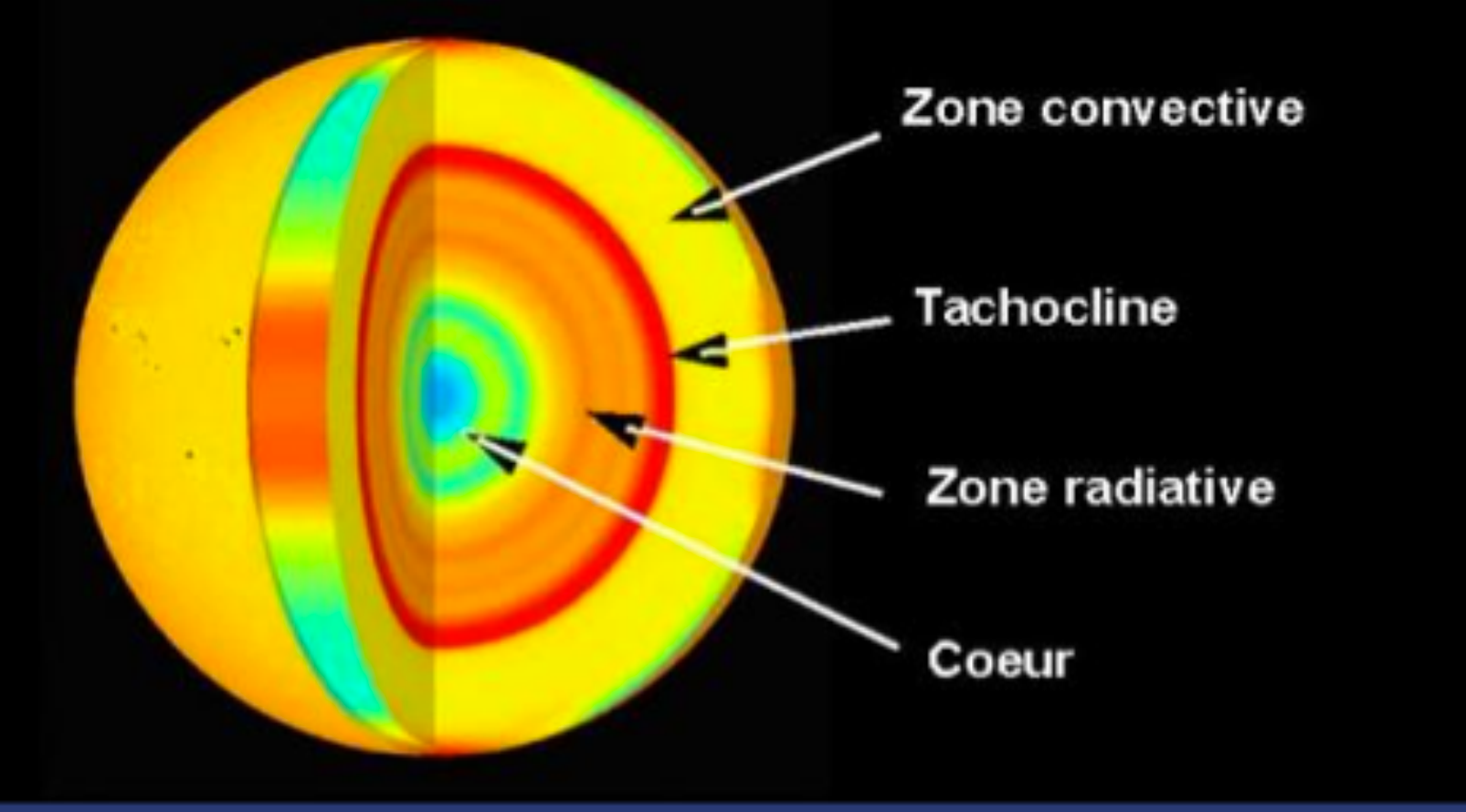}
\caption{\label{fig1} Composition of the Sun showing a cut along the radius of the difference between the sound-speed velocity computed between a model and the oscillation modes. The convective zone corresponds to the external 70\% of the radius. The rest of the Sun is the radiative interior. The tachocline is the transition region between the radiative and the convective zone. The core starts at the inner 30\% of the radius.
}
\end{figure}

Helio- and asteroseismology aim to study the internal structure and dynamics of the Sun and other stars by means of their resonant oscillations (e.g. Gough \cite{Gou1985}, Turck-Chize et al. \cite{stc93}; Christensen-Dalsgaard \cite{JCD2002}, and references therein). These vibrations manifest themselves in small motions of the visible surface of the star and in the associated small variations of stellar luminosity. 
During the last 30 years, helioseismology has proven its ability to study the structure and dynamics of the solar interior in a stratified way (see Figure~\ref{fig1}). These seismic tools allow us to infer some physical quantities as a function of the radius and latitude: the sound-speed (e.g. Basu et al. \cite{BasJCD1997}; Turck-Chi\`eze et al. \cite{STCBas1997}), the density profile (e.g. Basu et al. \cite{2009ApJ...699.1403B}), the internal rotation profile in the convective (e.g. Thompson et al. \cite{ThoToo1996}) and the radiative zone (Couvidat et al. \cite{CouGar2003}; Eff-Darwich et al. \cite{2008ApJ...679.1636E}; Garc\'\i a et al. \cite{GarCor2004,2008SoPh..251..119G}) or the conditions and properties of the solar core (e.g. Appourchaux et al. \cite{2010A&ARv..18..197A}; Garc\'\i a et al. \cite{2007Sci...316.1591G,2008AN....329..476G,2008SoPh..251..135G}; Turck-Chi\`eze et al. \cite{STCCou2001,STCGar2004}) are some well-known examples. Moreover, thanks to the detailed study of these variables, the position of the base of the convection zone (Christensen-Dalsgaard et al. \cite{JCDGou1985}) or the Helium abundances (Vorontsov et al. \cite{1991Natur.349...49V}) are some examples of what has been inferred. These observational constraints have significantly improved the standard solar models. 

\begin{figure}
\center
\includegraphics[height = 0.7\textheight]{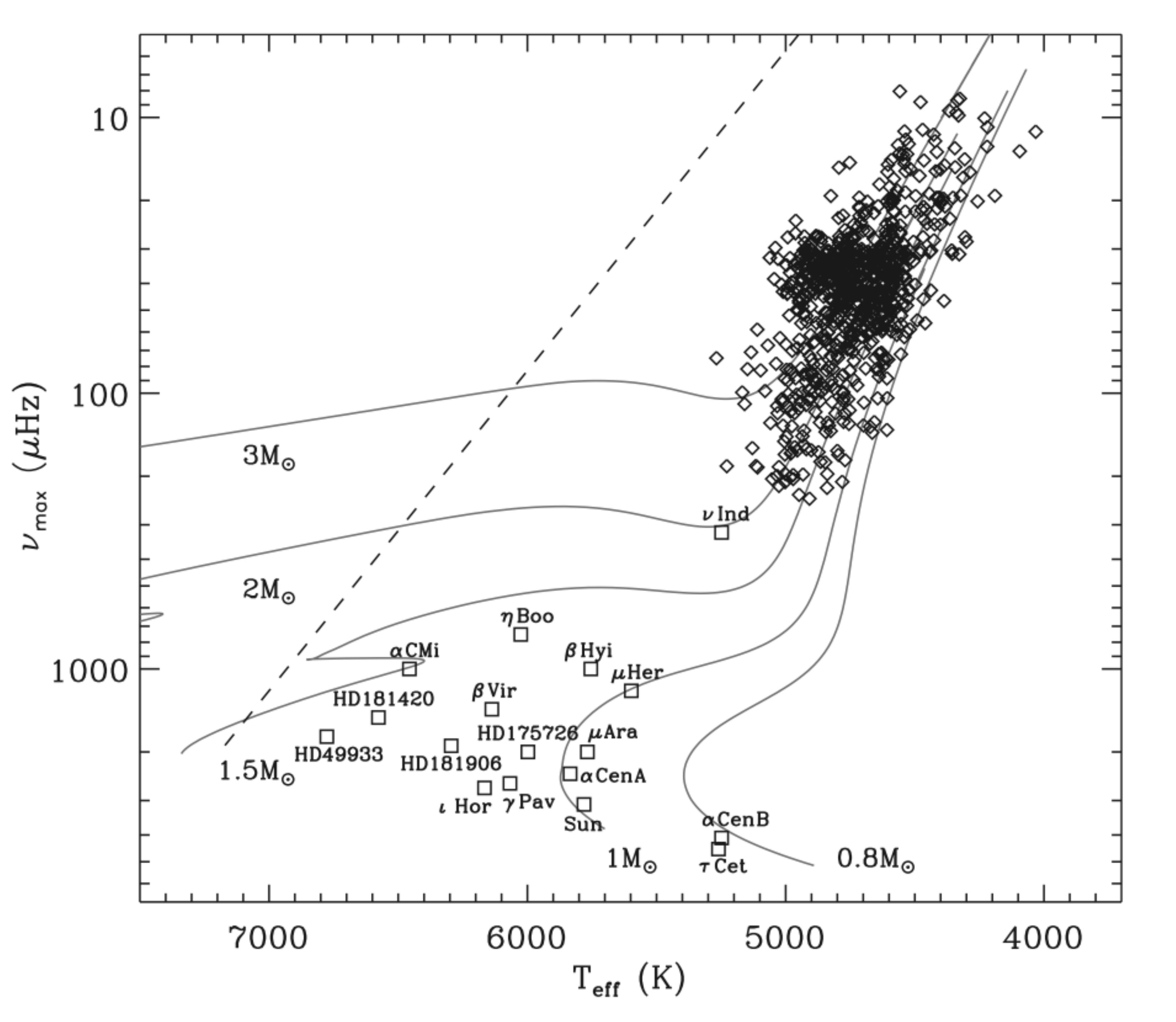}
\caption{\label{fig2} Seismic HR diagram (from Huber et al. \cite{2010ApJ...723.1607H}) in which the luminosity in the y-axis has been changed by the seismic variable $\nu_{max}$, i.e., the frequency of the maximum of the p-mode hump. The diamonds are 800 red giants measured continuously during several months {\it Kepler} scientific operations. The squares are main-sequence stars measured from ground- and space-based instruments (Stello et al. \cite{2009ApJ...700.1589S}). The continuos lines show solar-metallicity ASTEC evolutionary tracks (Christensen-Dalsgaard \cite{JCD2008}) for a range of masses. The dashed line marks the approximate position of the red edge of the instability strip.
}
\end{figure}

\begin{figure}
\center
\includegraphics[width = 0.98\textwidth]{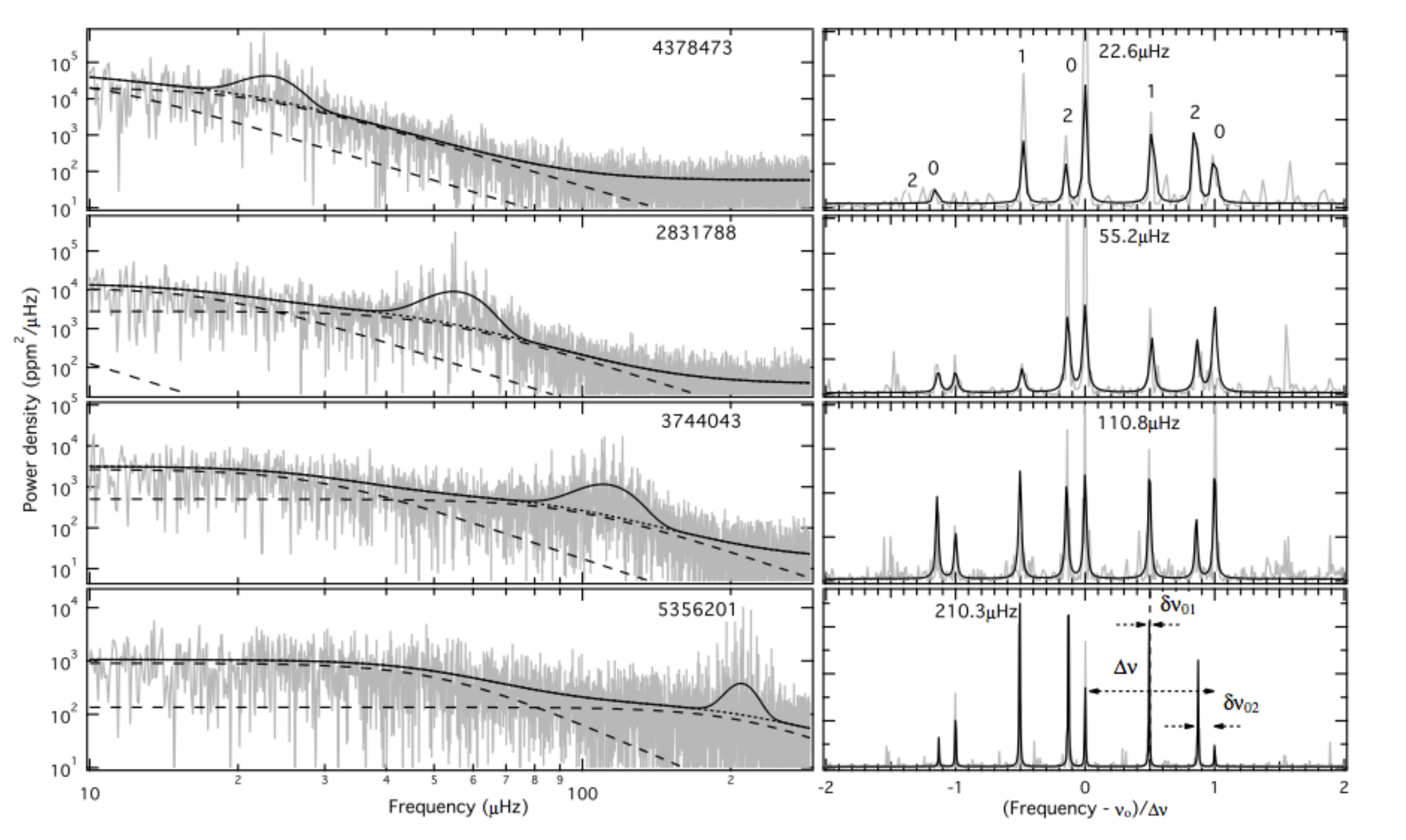}
\caption{\label{fig3} Example of the power spectra of 4 red-giant stars observed by {\it Kepler} (from Kallinger et al. \cite{2010A&A...522A...1K}). Left panels represent the full spectra in which the continuous lines are a global fit to the spectra while the dashed lines represent a model of the convective background. The right panels are the residuals around the p-mode humps once the background signal has been removed. Only low-degree modes ($\ell \le 3$) can be observed due to a geometrical cancellation of higher degree modes. The black lines correspond to the best fit of the individual frequencies. 
}
\end{figure}

The Sun, because of its proximity, has been a fundamental calibrator of stellar evolution but observations of many other stars Ñcovering the full HR diagram through asteroseismology (see Figure~\ref{fig2})Ñ will allow testing stellar evolution and dynamo theories under many different conditions (e.g. Christensen-Dalsgaard \& Houdek \cite{2010Ap&SS.328...51C}). Due to the absence of spatial resolution in the observations, only low-degree modes (those with a small number of nodal lines on the surface of the star, see Fig.~\ref{fig3})) will be accessible, therefore compared to the Sun, less detailed information will be available on stellar interiors. On the other hand, some pulsating solar-like stars offer the possibility to observe mixed (Arentoft et al. \cite{2008ApJ...687.1180A}; Bedding et al. \cite{2010ApJ...713..935B}; Chaplin et al. \cite{2010ApJ...713L.169C}; Deheuvels et al. \cite{2010A&A...515A..87D}) and maybe even gravity modes, thus to better constrain the structure and dynamics of their radiative interiors (e.g. Deheuvels et al. \cite{2010A&A...514A..31D}; Metcalfe et al. \cite{2010ApJ...723.1583M}).

Stars are also known to be magnetized rotating objects. Such dynamical factors, magnetism and rotation, affect the internal structure and evolution of stars (e.g. Brun et al. \cite{2004ApJ...614.1073B}; Zhan et al. \cite{2008sf2a.conf..341Z}; Duez et al. \cite{2010MNRAS.402..271D}; Eggenberger et al. \cite{2010A&A...519A.116E}). Thus, it is necessary to go beyond the classical modeling of stellar interiors and evolution by taking into account transport and mixing mechanisms both on dynamical and secular time-scales (e.g. Mathis \& Zhan \cite{2004A&A...425..229M,2005A&A...440..653M}; Turck-Chi\`eze et al. \cite{2010ApJ...715.1539T}).  Such models have improved our understanding of the history of rotation and magnetism in the entire HR diagram.  Moreover, the models are fundamental to understand observations such as the internal rotation profile in the solar radiative core (e.g. Mathur et al. \cite{2008A&A...484..517M}), the abundance anomalies observed on stellar surfaces (e.g. Mathur et al. \cite{2007ApJ...668..594M}, Zaatri et al. \cite{2007A&A...469.1145Z}), their magnetic fields (e.g. Jouve et al. \cite{2010A&A...509A..32J}).  However, theoretical and numerical models are assuming hypotheses that must be strongly constrained by observations. 

Nowadays, the increasing number of observations will also contribute to provide constraints for modelers by doing an extensive cartography of the rotation and convection of solar-like oscillating stars all along the HR diagram using {\it Kepler} data.

\section{The CoRoT mission}

The CNES-ESA mission CoRoT (Convection, Rotation and planetary Transits), launched on December 27, 2006, has been the first dedicated asteroseismic mission that has been able to perform ultra-high precision, wide-field, relative stellar photometry, for very long continuous observing runs (up to 150 days) on the same field of view. CoRoT has two main scientific programs working simultaneously on adjacent regions of the sky: asteroseismology, and the search for exoplanets.  Up to now, seven exoplanets have been discovered with CoRoT data and confirmed by ground-based follow-up campaigns. In February 2009, CoRoT discovered CoRoT-7b (L\'eger et al. \cite{2009A&A...506..287L}), one of the smallest exoplanets ever found. It is a rocky planet weighting 5 Earth mass orbiting around a solar-like star (G9V).  This was the first confirmed rocky exoplanet detected.  The accuracy of the CoRoT data is exemplified by the detection of the secondary transit of CoRoT-1b (Alonso et al. \cite{2009A&A...506..353A}), when the planet passes behind its star.  The comparison of the depths of both transits provides information on the albedo of the planet, hence on the nature of its atmosphere. 

In the asteroseismic context, CoRoT --after having operated during 4 years-- has already utterly modified our vision of pulsating stars. CoRoT has shown that the stellar oscillations are generally more complicated than those of the Sun. In the case of solar-like stars, more than 10 stars have already been observed. For 4 cases the determination of individual p-mode frequencies has reported  (Appourchaux et al. \cite{2008A&A...488..705A};
 Barban et al. \cite{2009A&A...506...51B}; Benomar et al. \cite{2009A&A...507L..13B}; Deheuvels et al. \cite{2010A&A...515A..87D}; Garc\'\i a et al. \cite{2009A&A...506...41G}) while in two other stars, the signal was too weak and only a power hump associated with p modes were detectable (Mathur et al. \cite{2010A&A...518A..53M}; Mosser et al. \cite{2009Mosser}). The rest of the stars are currently being analyzed. The predictions of excitation and damping mechanisms of acoustic modes based on the Sun observations are different from those inferred from the observations of the three first CoRoT solar-like targets (Michel et al. \cite{2008Sci...322..558M}).  

Concerning other classical pulsators, the first $\delta$-Scuti stars observed by CoRoT have shown thousands of excited modes while from Earth only a few dozens were detected (e.g. Poretti et al. \cite{2009AIPC.1170..435P}). Another discovery of CoRoT are hybrid stars showing different types of pulsations at the same time (e.g. Belkacem et al. \cite{2009Sci...324.1540B}). Finally, oscillations have been detected and quantified in hundreds of red giants (e.g. Mosser et al. \cite{2010A&A...517A..22M} and references there in) allowing as well, for the first time, the detection of non radial modes (de Ridder et al. \cite{deRidder09}) and some statistical studies using their seismic properties (Miglio et al. \cite{2009A&A...503L..21M}).  The physical processes responsible for the oscillations in these red giants are now better understood (e.g. Dupret et al. \cite{2009A&A...506...57D}). 

 In October 2009, the CoRoT operations were extended by 3 more years by the French space agency (CNES).

\section{The {\it Kepler Mission}}

Observational astroseismology reaches its maturity with the launch of {\it Kepler} on March 7, 2009 (GMT)  (Borucki et al. \cite{2010Sci...327..977B}; Koch et al. \cite{2010ApJ...713L..79K}). It is a NASA discovery mission whose primary goal is the search for and characterization of extrasolar planetary systems.  This will be accomplished by time-series photometry of around 150,000 stars in a single field of view of 115 $deg^2$ --selected to provide the optimal density of stars-- and located in the constellation of Cygnus for an initial mission lifetime of 3.5 years.  Planets transiting the visible disk of the star will be detected through the resulting dip in the light intensity; the magnitude of the dip provides a measure of the radius of the planet when the stellar radius is known.  The main scientific objective is to measure Earth-like planets in an Earth-like orbit around a Sun-like star inside the habitable zone.  The very precise photometry required for the planet search also provides excellent data for asteroseismology. While most stars will be observed at a cadence of 30 min, some of them (around 512 at any time), will be observed at a cadence of 1 min (Gilliland et al. \cite{2010ApJ...713L.160G}; Jenkins et al. \cite{2010ApJ...713L..87J}). Combining the stars observed at both cadences will allow us to study most of the stars in the HR diagram. 

The rich information content of these seismic signatures means that the fundamental stellar properties (e.g., mass, radius and age) may be measured to a precision not reached with any other classical method, which is fundamental to characterize the properties of the stars hosting planets and find the first telluric planet on the habitable zone, which is the main objective of {\it Kepler}. By combining the efforts of exoplanet researchers and asteroseismologists, it might be then possible to answer one of the most important open questions of mankind: are there any other planets in the universe with the conditions to have liquid water on their surface and develop life as it is known on Earth?

However, the {\it Kepler} Asteroseismic Investigation(KAI, Gilliland et al. \cite{2010PASP..122..131G}; Kjeldsen et al. \cite{2010arXiv1007.1816K}) has as a much larger objective: to investigate stellar structure and evolution over a broad selection of stars. To accomplish this goal, an agreement has been established between NASA and an international consortium coordinated by the Danish group at the University of Aarhus. First, this involves making the asteroseismic relevant data available to the entire {\it Kepler} Asteroseismic Science Consortium (KASC) (http://astro.phys.au.dk/KASC/) currently involving more than 400 scientists. The activities within KASC include development of data pipelines for an initial analysis to characterize the global properties of the stars, particularly their radii and ages (examples of these pipelines, in the case of solar-like stars can be found in: Campante et al. \cite{2010MNRAS.408..542C}; Hekker et al. \cite{2010MNRAS.402.2049H}; Huber et al. \cite{2009CoAst.160...74H}; Mathur et al. \cite{2010A&A...511A..46M}; Mosser \& ppourchaux \cite{2009A&A...508..877M}).Then it consists of developing and later applying more sophisticated techniques for data analysis and interpretation. Finally, it would involve further development of stellar modeling techniques, in anticipation of, and response to the {\it Kepler} results to obtain a deeper understanding of the physical processes in stellar interiors (e.g. Metcalfe et al. \cite{2010ApJ...723.1583M}). 

The first results of the initial observations done by {\it Kepler} are very promising. The analysis of the first three short-cadence, 1-month long, solar-like stars showed an acoustic power spectrum of unprecedented quality (Chaplin et al. \cite{2010ApJ...713L.169C}) while the study of many long-cadence G and K giants extending in luminosity from the red clump down to the bottom of the giant branch showed solar-like oscillations with degree ² 3 (Bedding et al. \cite{2010ApJ...713L.176B}) confirming some theoretical scaling laws (Hekker et al. \cite{2010arXiv1008.2959H}; Huber et al. \cite{2010ApJ...723.1607H}) and confirming fundamental stellar parameters (Kallinger et al. \cite{2010A&A...522A...1K}). Also, 47 stars in the field of the open cluster NGC 6819 were analyzed by Stello et al. \cite{2010ApJ...713L.182S} showing that asteroseismic parameters allow us to test cluster-membership of the stars, and even with the limited seismic data in hand, it was possible to identify four potential non-members despite the fact that they have a better than 80\% membership probability from radial velocity measurements.

Concerning the asteroseismic study of solar-like stars by {\it Kepler}, a survey phase has been done during the first 10 months of {\it Kepler} nominal operations in which 2000 solar-like stars have been observed with a 1 minute cadence (Chaplin et al. in preparation). The main goal of this phase is to provide the data that will allow us to select the best solar-like stars (around a hundred) to be targeted during the rest of the nominal mission for, at least, two and a half years, and to cover most of the HR diagram with a sufficiently high number of stars to be able to do some comparative studies with good-enough statistics. During the survey phase each star has been observed for one month at a time, putting constraints on the way to extract the seismic parameters. These data are currently being analyzed. For the brightest targets we will be able to get individual frequencies (a few hundreds), but for the majority of them (fainter ones) this will not be the case and one must rely on extracting mean global seismic parameters.

\section{Conclusions}

Stars are building blocks of the universe. They are responsible for the chemical evolution of the galaxy and they are crucial sources of energy for giving conditions which can support life. Stellar evolution theory feeds into stellar, galactic, extra-galactic and cosmology studies. Thus, any improvement in stellar modeling will have a direct impact in all the above mentioned domains. 

Asteroseismology is in a position to provide new constraints to the physical properties of the stars coupled to a  detailed knowledge of their structure and dynamics. It provides accurate determination of the masses, radii and ages of stars, to a level not reached by any other classical methodology. It can also be used to measure the enrichment levels of helium in solar-type stars with a big implication for cosmology.

Finally, seismology is playing a major roll in the exoplanet research and the study of habitable zones.

\section*{Acknowledgments}   
%
CoRoT (Convection, Rotation and planetary Transits) is a mini-satellite developed by the French Space agency CNES in collaboration with the Science Programs of ESA, Austria, Belgium, Brazil, Germany and Spain. Funding for the {\it Kepler} Discovery mission is provided by NASAÕs Science Mission Directorate. This work has been supported by the French PNPS program and the IRFU/SAP. The author also wants to thank the support providing by SEA.

%

%
\end{document}